\documentclass{rstransa}

\begin{document}

\title{Collisions involving antiprotons and antihydrogen: an overview}

\author{
S. Jonsell}

\address{Department of Physics, Stockholm University, SE-10691 Stockholm, Sweden.}

\subject{Atomic and Molecular Physics}


\corres{Svante Jonsell\\
\email{jonsell@fysik.su.se}}

\begin{abstract}
I give an overview of  experimental and theoretical results for  antiproton and antihydrogen scattering with atoms and molecules (in particular H, He). At low energies ($\lesssim1$~keV) there are practically no experimental data available. Instead I compare the results from different theoretical calculations, of various degrees of sophistication. At energies up to a few 10:s of eV,  I focus on simple approximations that give reasonably accurate results, as these allow quick estimates of collision rates without embarking on a research project.
\end{abstract}


\begin{fmtext}
\section{Introduction}

 There are several reasons why collisions between antiprotons or antihydrogen and ordinary atoms or molecules are of interest to current and future experiments. Most obvious is, perhaps, that the interaction with background gas limits the lifetime of trapped antiparticles and antiatoms. For this purpose, knowledge of the relevant cross sections can guide experiments, \emph{e.g.}, which gases will be most problematic, which loss processes will dominate in different energy ranges, are there resonances which will cause enhanced losses for certain species and energies, what can we expect for the relative lifetimes of different antimatter species?

However, effects from matter-antimatter collisions are not only negative, there are also instances when matter-antimatter collisions are introduced on purpose.
In ASACUSA antiprotons are stopped in helium, loosing their energy through ionising collisions, followed by antiproton capture into antiprotonic helium \cite{Hayano07}. In the future one may want to use similar reactions to create other forms of mixed matter-antimatter states, \emph{e.g.}, the H$\overline{\mathrm{H}}$ molecule. Collisions between ${\rm H}_{2}^+$ ions 
and antiprotons have also led to the formation of protonium, \emph{i.e.}, the bound state between a proton and an antiproton \cite{Zurlo06}.
Another suggestion is to sympathetically cool antihydrogen using trapped and cooled atoms.  
\end{fmtext} \maketitle

Matter-antimatter interaction is also interesting in its own right. This is a new branch of atomic physics, which faces completely new  challenges, not only experimentally but also theoretically. In particular, capture processes leading to the formation of positronium (Ps) or protonium (Pn) is a challenge for theory.

Below, I shall start with antiprotons colliding with H and He at energies on the keV scale, and thereafter progress to lower energies. 
As the collision energy is reduced more channels, as well as resonant structures, become important. This makes it very difficult to obtain accurate numerical results, and only few theoretical results available at energies below a few 10:s of eV. I will therefore focus on  simple ways to obtain estimates of the relevant cross sections, and compare these to more exact calculations when available.
Then I will discuss antihydrogen-atom interactions at low energies.

\section{Antiprotons}

\subsection{Ionisation of atoms and molecules by high-energy antiprotons}

 In this overview, high energies refers to $\sim 1$~keV and above, even though in other contexts this is referred to as low energies.
This is the most thoroughly researched area, and essentially the only regime where there are experimental data available. Experimental and theoretical developments was reviewed in detail by Kirchner and Knudsen in 2011 \cite{KK2011}, here I will only summarise a small part of this work. Since this reviewed appeared there has also been a lot of progress using the convergent close-coupling (CCC) method by Abdurakhmanov and co-workers \cite{Abdurakhmanov2011a,Abdurakhmanov2011b,Abdurakhmanov2013,Abdurakhmanov2014,Abdurakhmanov2015,Abdurakhmanov2017}.

 The first experiments were done at LEAR in the second half of the 1980:s \cite{Andersen86}, and have been continued into the present by the ASACUSA collaboration at the Antiproton Decelerator (AD). Many targets have been studied: H, D, He, Ne, Ar, Kr, Xe, H$_{2}$, D$_{2}$, N$_{2}$, CO, CO$_{2}$, CH$_{4}$, O$_{2}$ at energies ranging from 3~keV to 20~MeV.  Antiprotons are useful as probes of atomic and molecular structure, since the collisions are not complicated by electron transfer (unlike positively charged projectiles) or exchange processes (unlike electrons). Thus antiprotons provide a very clean ionisation signal. Charge effects can be studied by comparing to proton scattering.

\subsection{Antiproton--hydrogen}

At energies \emph{above} 200~keV the experimental results \cite{Knudsen95} are well described by the first-Born approximation. As this approximation is based on first-order perturbation theory, the cross section is independent of the charge of the projectile. Indeed, the measured ionisation cross section is identical to that of protons, within the uncertainty of the measurement.

For energies \emph{below} 200~keV, the charge of the projectile does play a role. Since the proton attracts the electron, while the antiproton repels it, the proton cross section becomes larger than the antiproton cross section. In fact, while the proton cross section monotonically increases with decreasing collision energy, the antiproton cross section levels out, and even declines slightly, below collision energies around 150~keV. Down to about 50~keV the experimental results are well described by the distorted wave Born approximation. 

Below 30~keV there are no experimental data, so instead I compare the results from different theoretical approaches. Theoretical calculations in this regime are usually based on classical motion of the antiproton, creating a time-dependent perturbation of the hydrogen atom. I will refer to this as the \emph{semi-classical} method. This method comes in two varieties, either the antiproton is assumed to follow a completely straight trajectory, unperturbed by the interaction with the atom, or the bending of the antiproton trajectory is allowed for. The latter method treats only the antiproton--atom separation $R$ classically, while the angular variables are treated quantum mechanically, leading to rigorous conservation of quantum mechanical angular momentum \cite{Sakimoto00}.  The time-dependent Schr\"odinger equation for the hydrogen atom is then solved using various methods. Thus, if the solution to the quantum problem is fully converged, the different calculations should agree. Down to about 1~keV the agreement is about 10-15\%, showing that at least some calculations have convergence problems. As the energy is reduced by another order of magnitude, the spread between different calculations increases to about 50\%. However, in this regime it seems to be important to allow for the bending of antiproton trajectories; calculations doing this show a much better agreement.

\begin{figure}[!h]
\centering\includegraphics[width=10 cm]{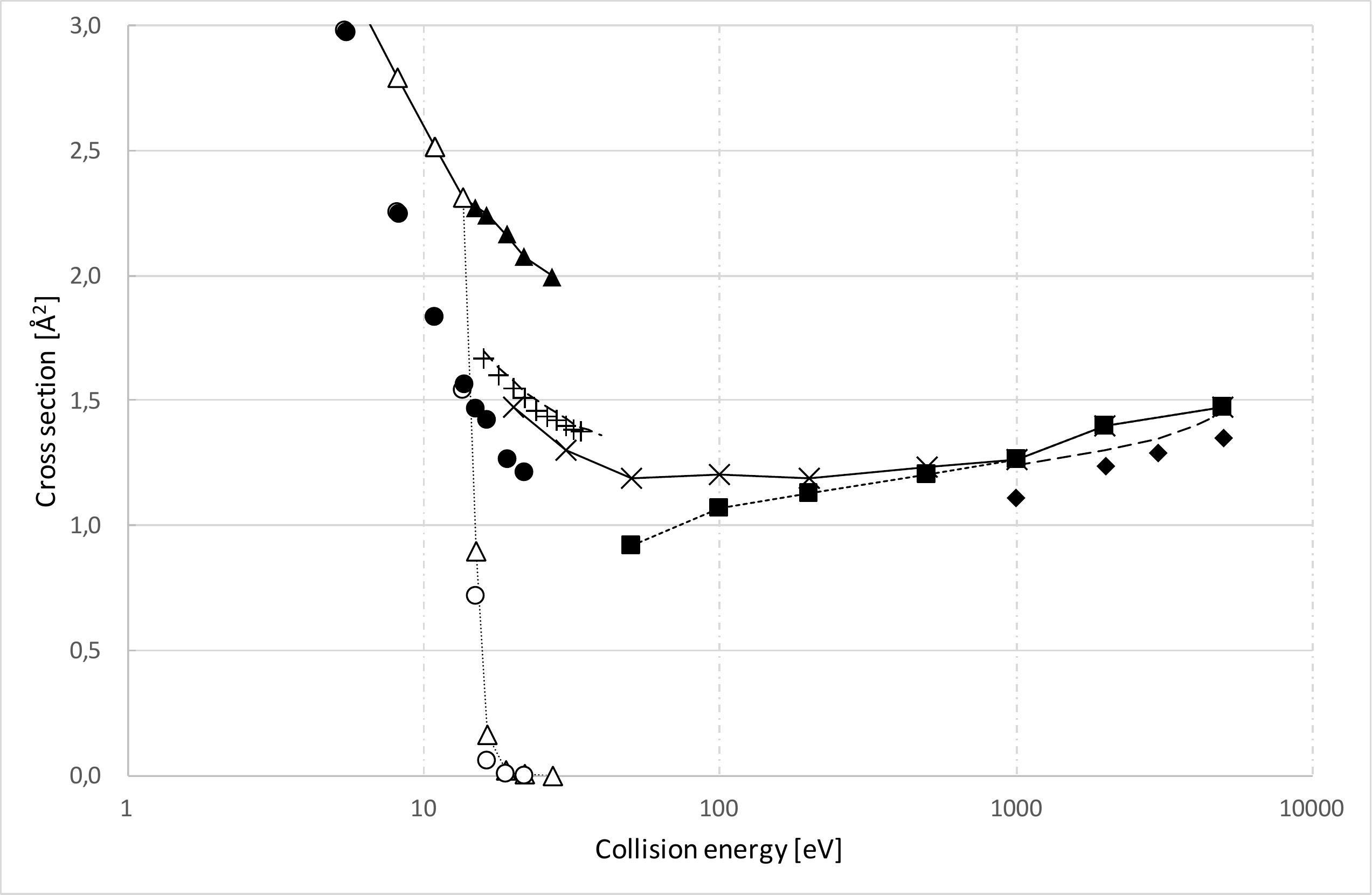}
\caption{Cross sections for ionisation and protonium formation in antiproton collisions with hydrogen, calculated using different methods: \emph{filled diamonds}, ionisation fully quantum \cite{Abdurakhmanov2011a}; \emph{filled squares}, ionisation calculated semi-classically using straight line antiproton trajectories \cite{Sakimoto00}; \emph{crosses}, the same but allowing for bending of the trajectories  \cite{Sakimoto00};  \emph{long dashes}, semi-classical allowing for bent trajectories using a Laguerre mesh \cite{Sakimoto00b,Sakimoto04}; \emph{plusses}, ionisation, fully quantum \cite{Sakimoto04};\emph{filled triangles}, total cross section calculated with CTMC; \emph{open triangles}, Pn formation calculated with CTMC; \emph{filled circles}, total cross section using FMD; \emph{open circles}, Pn formation using FMD. All results using CTMC and FMD are taken from \cite{Cohen97}.  }
\label{fig:Hion}
\end{figure}

Figure \ref{fig:Hion} shows the situation in the region 5~eV to 5~keV. Note that the semi-classical calculations using straight and bending antiproton trajectories agree for energies above 1~keV, while at lower energies they have different trends. However, for energies $>1$~keV both give $\sim10$ \% larger cross sections than the fully quantum mechanical results by Abdurakhmanov \emph{et al.} \cite{Abdurakhmanov2011a}. As expected, the calculation allowing for bending agrees much better with the presumably accurate fully quantum calculations, especially with the improved numerical treatment using a Laguerre mesh \cite{Sakimoto00b,Sakimoto04}. Below 13.6~eV, the antiproton does not have enough energy to ionise the atom, so the ionisation cross section drops to zero. However, the curves for the \emph{total} inelastic cross section (ionisation + protonium formation) cross the ionisation threshold rather smoothly. Note that above the ionisation threshold protonium formation drops to zero very quickly. Thus, above 20 eV the total cross section is equal to the ionisation cross section.

 In the traditional Classical Trajectory Monte Carlo (CTMC) method, particle trajectories are calculated using purely classical equations of motion, with randomised initial conditions.
Below and around the ionisation threshold, the CTMC calculations lie about 30--40\% above the fully quantum mechanical results.

A method similar to CTMC, the Fermion Molecular Dynamics (FMD) method \cite{Cohen95,Cohen97,Cohen04}, seems to do much better, even though its results are some 15\% too low. The difference is that a "Heisenberg potential", with the purpose to mimic the Heisenberg uncertainty principle, is added to the equations of motion. This potential has the form
\begin{equation}
V_H(p,r)=\frac{\hbar^2}{2mr^2}\frac{\xi^2}{4\alpha}\exp\left\{\alpha\left[1-\left(\frac{rp}{\xi\hbar}\right)^4\right]\right\},
\label{eq:VH}
\end{equation}
where $r$ and $p$ are the  position and momentum of the electron relative the proton, $\alpha=4$ and $\xi$ is tuned to reproduce the true ionisation energy of the hydrogen atom. Since the potential is momentum dependent, Hamilton's equations of motion, rather than Newton's,  have to be solved. When $rp>\xi\hbar$  the potential rapidly goes to zero. In the opposite limit it creates a repulsive barrier, so that $r$ and $p$ cannot simultaneously become small. When applied to many-electron atoms $V_H$ stabilises the atoms against autoionisation, which will occur if a purely classical description is used. While not necessary to stabilise the hydrogen atom, this addition evidently improves the accuracy of the calculations.

\begin{figure}[!h]
\centering\includegraphics[width=10 cm]{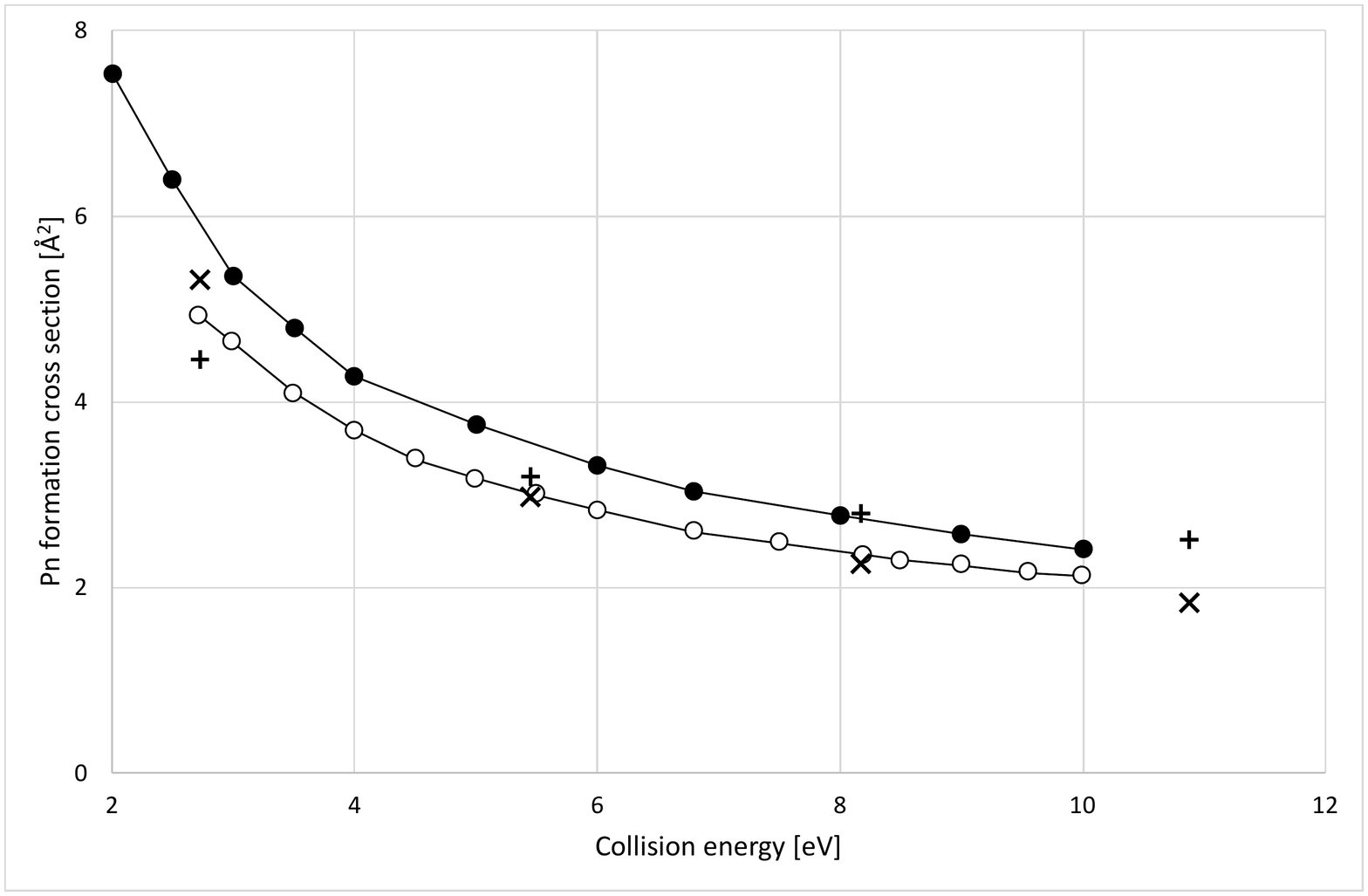}
\caption{Cross sections for protonium formation  in collisions between antiprotons and hydrogen atoms, for collision energies below the ionisation threshold, calculated using different methods: \emph{open circles}, fully quantum \cite{Sakimoto01PRA}; \emph{filled circles}, semi-classical allowing for bending of the antiproton trajectory  \cite{Sakimoto01jpb}; \emph{plusses}, CTMC \cite{Cohen97}; \emph{crosses}, FMD  \cite{Cohen97}.  }
\label{fig:Pnformation}
\end{figure}

In figure \ref{fig:Pnformation}, we take a closer look at the protonium formation cross section in the range 2--12 eV. As there are no experimental data, I take the fully quantum mechanical  calculation by Sakimoto  \cite{Sakimoto01PRA} as my reference. The semi-classical results \cite{Sakimoto01jpb} show the same trend as the quantum results, but are about 15\% too high. The FMD results \cite{Cohen97} also have roughly the right trend, and actually seem to agree better with the fully quantum results, though this is likely to be coincidental. The CTMC results \cite{Cohen97} also lie close to the others, but the curve is much flatter. Considering that the quantum calculations are very complex, and unlikely to be practical for heavier atoms, these results suggest that the much simpler FMD method is a viable option at eV-scale energies.

Moving to even lower energies, the behaviour of the cross sections can be understood by considering the angular momentum barrier. For a collision with impact parameter $b$ and collision energy $E$, the angular momentum is $L=mvb=\sqrt{2mE}b$, and the effective potential becomes
\begin{equation}
V_{\rm eff}(R)=V_{\mathrm{ad}}(R)+E\frac{b^2}{R^2},
\label{eq:Veff}
\end{equation}
where $V_\mathrm{ad}$ is the adiabatic potential between the antiproton and the atom. For the interaction between a charged particle and an atom, the long-range form of the potential is attractive
\begin{equation}
V_\mathrm{ad}\xrightarrow{R\rightarrow\infty}-\frac{e^2}{4\pi\epsilon_0}\frac{\alpha}{2R^4},
\label{eq:Vpol}
\end{equation}
where $\alpha$ is the polarisability of the atom. A very simple, but often quite accurate, approximation to the total inelastic cross section can be calculated by assuming that if the projectile has energy enough to overcome the angular momentum barrier, a reaction occurs with 100\% probability, but if it doesn't no reaction can occur. If $b$ is sufficiently small, the barrier will not block antiprotons of any energy. Thus for a given collision energy $E$ there is a maximum impact parameter $b_c(E)$, for which  inelastic collisions are possible. The total inelastic cross section is then
\begin{equation}
\sigma(E)=\pi b_c(E)^2.
\label{eq:sigma}
\end{equation}
 At sufficiently low $E$, the form  (\ref{eq:Vpol}) will be valid at the classical turning point, which for $b=b_c(E)$ is the maximum of $V_\mathrm{eff}$. It is then easy to derive an analytical form for the cross section,
 \begin{equation}
 \sigma_L=\pi\sqrt{\frac{e^2}{4\pi\epsilon_0}\frac{2\alpha}{E}}.
 \label{eq:Langevin}
 \end{equation}
 This is called the Langevin cross section. A lower limit of validity is set by the energy where this type of semi-classical description cannot be used. That is, we require $L=mvb_c(E)\gg\hbar$, which for an antiproton translates to $E\gg1\mu$eV$=10$mK. An upper limit of validity is set by the requirement that the maximum of the angular momentum barrier should lie within the region where (\ref{eq:Vpol}) is valid. This gives the condition $E\lesssim \alpha e^2/(8\pi\epsilon_0 R_*^4)$, where $R_*$ is the antiproton-nucleus separation where (\ref{eq:Vpol}) starts to break down. For the $\bar{p}$--H system $R_*\sim 3a_0$, giving $E\lesssim 0.7$ eV.

 If the entire adiabatic potential is known, then the this method can be extended to higher energies,  provided that the atom would emit an electron if the nuclear charge would be reduced by one unit (\emph{i.e.}, atomic numbers $Z$, where $Z-1$ does not have a stable negative ion). The adiabatic potential of such a system will merge with the electronic continuum at some finite $R$. For the $\bar{p}$--H system, this distance is given by the minimum dipole moment which binds an electron. This distance was determined to $R_{\mathrm{FT}}=0.639a_0$ by Fermi and Teller \cite{FermiTeller}.
If the antiproton reaches  $R<R_{\mathrm{FT}}$ it will be captured by the nucleus, while the atom emits an electron. For impact parameters and energies where there is a classical turning point outside $R_{\mathrm{FT}}$, one assumes that there is no antiproton capture. Thus, in the same way as for the Langevin form, we can use the potential to derive a simple form for the cross section, even though no general analytical expression can be given in this case.
 I call this the adiabatic cross section $\sigma_\mathrm{ad}$. Clearly, $\sigma_\mathrm{ad}$ should converge to $\sigma_L$ at low energies. 
 
 \begin{figure}[!h]
\centering\includegraphics[width=10 cm]{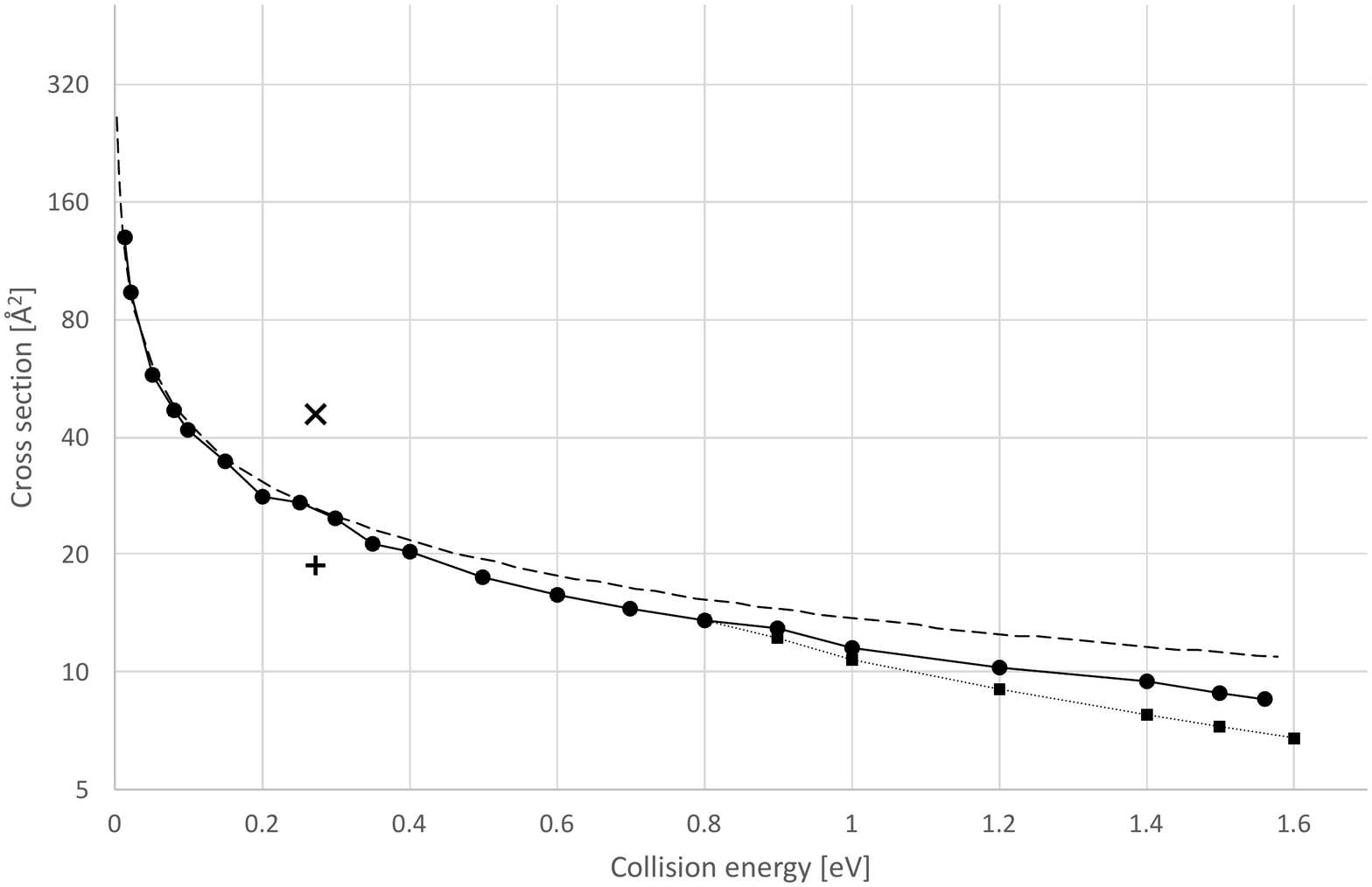}
\caption{Low-energy cross section for Pn formation in collisions between antiprotons and hydrogen atoms, calculated using different methods: \emph{filled circles},  semi-classical \cite{Sakimoto01jpb}; \emph{filled squares}, adiabatic  \cite{Sakimoto01jpb}; \emph{dashed line}, Langevin; \emph{cross}, FMD \cite{Cohen97}; \emph{plus}, CTMC  \cite{Cohen97}.  }
\label{fig:LE}
\end{figure}

 In figure \ref{fig:LE} I compare the Langevin and adiabatic cross sections to a semi-classical calculation \cite{Sakimoto01jpb}. Unfortunately, I am not aware of any fully quantum mechanical study in this energy regime. We find that the adiabatic and the Langevin cross sections agree quite well with the semi-classical results. These can also be regarded as semi-classical methods as atomic properties are calculated quantum mechanically (the interaction potential for the adiabatic method, the polarisability for the Langevin method). The difference between the semi-classical method of \cite{Sakimoto01jpb} and the adiabatic method, is that in the former approach the capture probability is calculated for each $R$, while  the latter approach assumes a step from 100\% for $R<R_{\mathrm{FT}}$ to 0 for  $R>R_{\mathrm{FT}}$. Since the adiabatic cross section is lower than the semi-classical, the adiabatic method must underestimate the capture probability for  $R>R_{\mathrm{FT}}$. Indeed, just outside  $R_{\mathrm{FT}}$ the energy gap to the continuum is very small, and hence non-adiabatic transitions are likely to give a significant contribution. At the lowest energies the Langevin cross section agrees perfectly with the semi-classical result, though we should keep in mind that we do not know how accurate the latter is. In contrast, both the CTMC and FMD methods, based on classical representations of the hydrogen atom, do not give good results.

\subsection{Antiproton-helium collisions}

Also for this process, experimental data are only available in the keV-regime (above 3~keV).  The single ionisation cross section shows features similar to the hydrogen case. For helium it is also possible to get double ionisation. Interestingly, while the single ionisation does not depend on the charge of the projectile at high energies, double ionisation differs between antiprotons and protons, even at energies as high as 1 MeV. The reason is that double ionisation is the sum of two processes, with different dependencies on the charge $q$ of the projectile. First, the projectile may by itself knock out two electrons. The matrix element for this process is proportional to $q^2$, since the projectile interacts twice. Second, the projectile may knock out one electron, which  on its way out in turn knocks out a second electron, or the second electron is emitted in a shake-off process by the rapid change of the potential in which it moves. The matrix element for this process is proportional to $q$, since the projectile only interacts  once. As long as both processes are present with similar amplitudes, the cross section is proportional to the square of the sum of the matrix elements, yielding an interference term proportional to $q^3$, \emph{i.e.}, depending on the charge of the projectile. This explains the difference between double ionisation from collisions with protons and antiprotons \cite{KK2011}.

For ionisation of helium, basically the same theoretical methods as for scattering on hydrogen  are available. However, implementing  them is considerably more difficult, as the correlation between the two electrons is difficult to capture accurately numerically. Scattering calculations for helium, and heavier  atoms, therefore often employ some effective one-electron model for the atom, or at least neglect correlation. This works reasonably well for single ionisation at energies above 50~keV \cite{KK2011}, but does not work for double ionisation.

At lower energies there is an additional problem with the one-active-electron model.  The  attraction between the antiproton and the nucleus will make it possible for the antiproton to penetrate also the inner electron core, even in the limit of zero collision energy. When this happens \emph{all} of the electrons in the atom will be strongly perturbed. 
\begin{figure}[!htb]
\centering\includegraphics[width=10 cm]{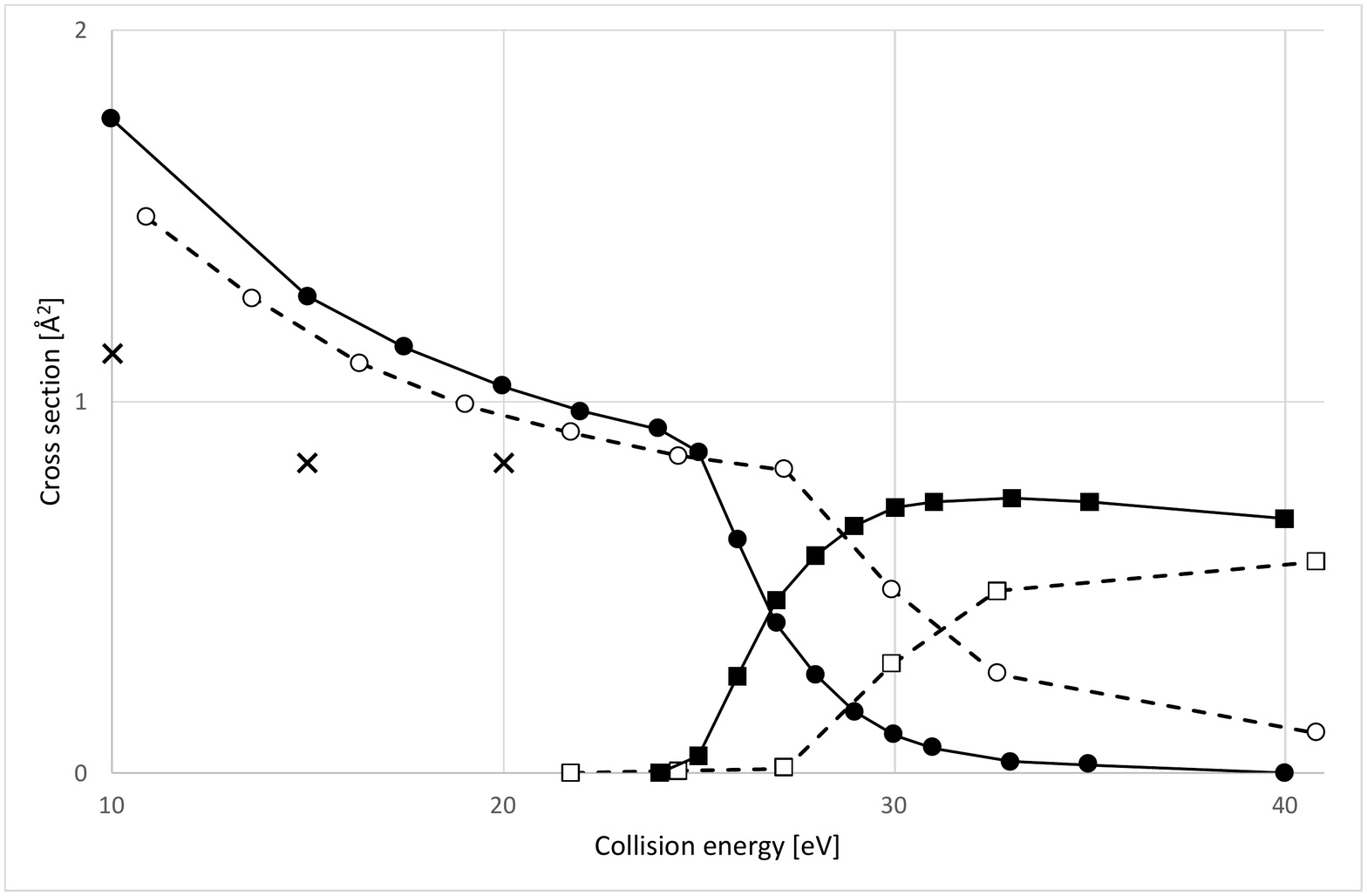}
\caption{Cross sections for antiproton collisions with helium: \emph{circles}; antiproton capture, \emph{squares}; ionisation. The filled symbols show calculations using the semi-classical method \cite{Sakimoto15}, the open symbols show FMD  \cite{Cohen00}. The crosses show antiproton capture from a  quantum calculation by Tong \emph{et al.} \cite{Tong08}. Note, the
results in \cite{Sakimoto15} have been multiplied by a factor 2 as suggested in that publication. }
\label{fig:He}
\end{figure}

In figure \ref{fig:He} I compare the semi-classical one-active-electron calculations of antiproton capture and of ionisation \cite{Sakimoto15} to FMD calculations of the same processes \cite{Cohen00}. As suggested by the author, the results in \cite{Sakimoto15} were multiplied by 2 to account for the fact that either of the electrons could participate in the collision. For the capture process, there are also results from a  quantum mechanical calculation, also in the one-active-electron approximation by Tong \emph{et al.} \cite{Tong08}. While methods based on a classical description of the atom, such as FMD, of course do not give a fully realistic description of the atomic structure, they have the advantage that they do include all particles on the same footing. Thus all correlations are included, albeit only on a classical level.

Below the ionisation threshold (24.6 eV), the FMD and semi-classical results agree fairly well. Unfortunately though, the quantum calculation which, at least in principle, should be more accurate than the semi-classical, gives a significantly smaller cross section. Just above the ionisation threshold the FMD and semi-classical approach differ significantly. This is not surprising since the final state of the reaction involves a low-energy electron, which should not be well described classically. At higher energies, the results of the two methods again seem to approach each other.

\begin{figure}[!h]
\centering\includegraphics[width=10 cm]{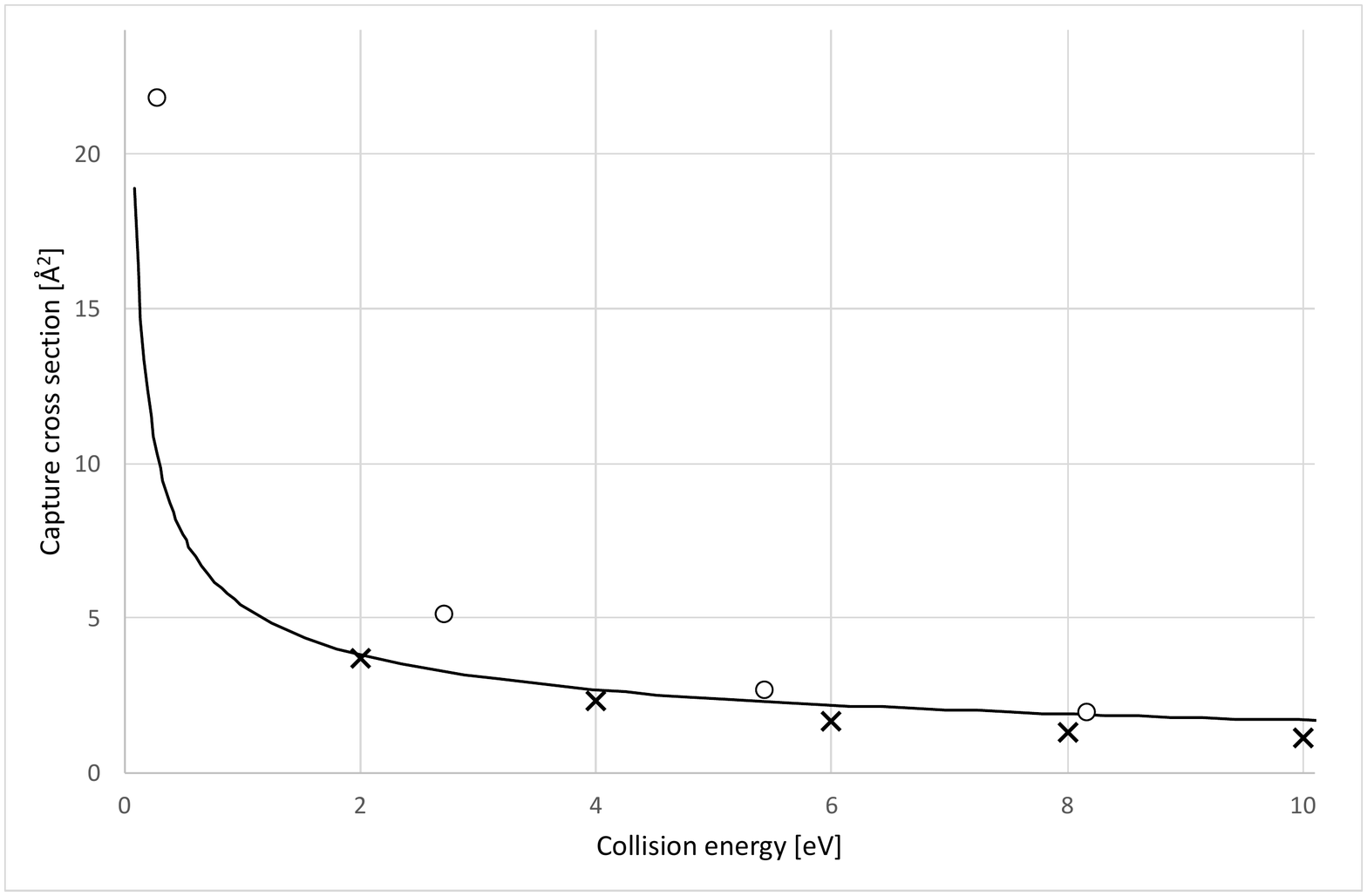}
\caption{Low-energy cross sections for antiproton capture in collisions with helium: \emph{solid line}, Langevin; \emph{crosses},  Tong \emph{et al.}  (one-active-electron quantum) \cite{Tong08}; \emph{open circles}, FMD \cite{Cohen00}. }
\label{fig:HeLE}
\end{figure}

Finally, I compare the calculation by Tong \emph{et al.} \cite{Tong08} to FMD  \cite{Cohen00} and Langevin cross sections at energies below 10 eV (figure \ref{fig:HeLE}). The agreement is relatively good, considering the crudeness of the latter methods, though this may very well be coincidental. Note that this is really above the energy regime where the Langevin cross section should be accurate. Since the H$^-$ ion is stable, the adiabatic curve for the helium-antiproton interaction never merges with the continuum. As a result, the adiabatic cross section is not defined. An alternative approach based on diabatic states is described in \cite{Cohen04}.

\subsection{Antiproton--hydrogen molecular ion}

Formation of protonium through collisions between antiprotons and H$_2^+$ ions was detected by the ATHENA experiment \cite{Zurlo06}. Two calculations exist in this regime: FMD \cite{Cohen05}, and a semi-classical calculation based on classical motion on the Born-Oppenheimer surface \cite{Sakimoto04H2+}. The results from the two calculations agree surprisingly well for both protonium formation and ionisation, over the energy range 10--50 eV, though there are some differences below 10 eV.

At higher energies ($>500$ eV), there are two quantum mechanical calculations for ionisation  of the ion \cite{Sakimoto05,Luhr09}. The two calculations are in perfect agreement.  The results in \cite{Luhr09} also include excitation of the ion.

\section{Antihydrogen collisions}

\subsection{Antihydrogen--hydrogen atom}

\begin{figure}[!h]
\centering\includegraphics[width=10 cm]{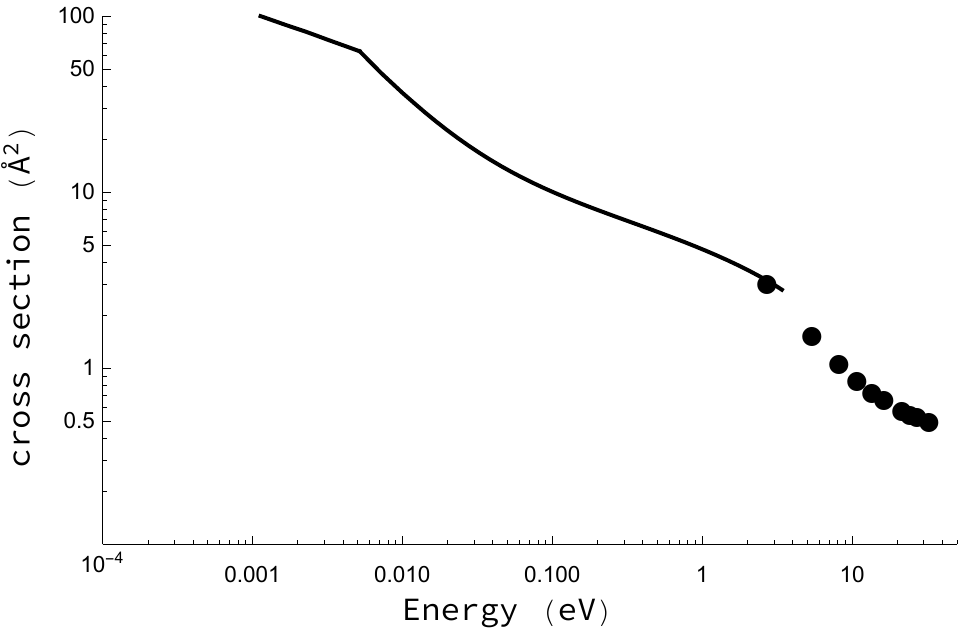}
\caption{Cross section for inelastic hydrogen--antihydrogen scattering: \emph{solid line}, adiabatic; \emph{filled circles}, FMD \cite{Cohen06}. }
\label{fig:hhbar}
\end{figure}

For antihydrogen experiments, the relevant energy scale is from a few eV, down to a few mK ($\sim\mu$eV). Thus, the focus is on much lower energies, compared to antiproton scattering. As the projectile has an internal quantum structure, the semi-classical method described above cannot be used. Furthermore, the Langevin formula does not apply to scattering between neutrals. However, similar considerations applied to the long-range van der Waals form $V(R)\rightarrow -C_6/R^6$, gives the cross section
\begin{equation}
\sigma(E)=3\pi\left(\frac{C_6}{4E}\right)^{1/3}.
\label{eq:sigC6}
\end{equation}
Unfortunately, due to the weaker long-range attraction, this form has a much more limited range of validity than the Langevin form. 
However, we can still use the adiabatic method, provided that adiabatic potential is known, and has a Fermi-Teller radius where it merges with the continuum (in this case, leading to  emission of ground-state positronium). For the hydrogen-antihydrogen system $R_{\mathrm{FT}}\lesssim 0.744a_0$  \cite{Strasburger02jpb}. The adiabatic cross section, calculated using the very accurate potential from  \cite{Strasburger02jpb}, is displayed in figure \ref{fig:hhbar}, along with FMD results at higher energies \cite{Cohen06}. Though the energy regimes are different, the two methods connect at 2 eV, where the results agree to within 5\%. The kink in the adiabatic cross section corresponds to a second maximum at smaller $R$ taking over as the global maximum for $E\gtrsim 0.005$ eV. This energy sets the upper limit of approximate validity of   eq.\ (\ref{eq:sigC6}), which just below is about 15\% smaller. 

\begin{figure}[!h]
\centering\includegraphics[width=10 cm]{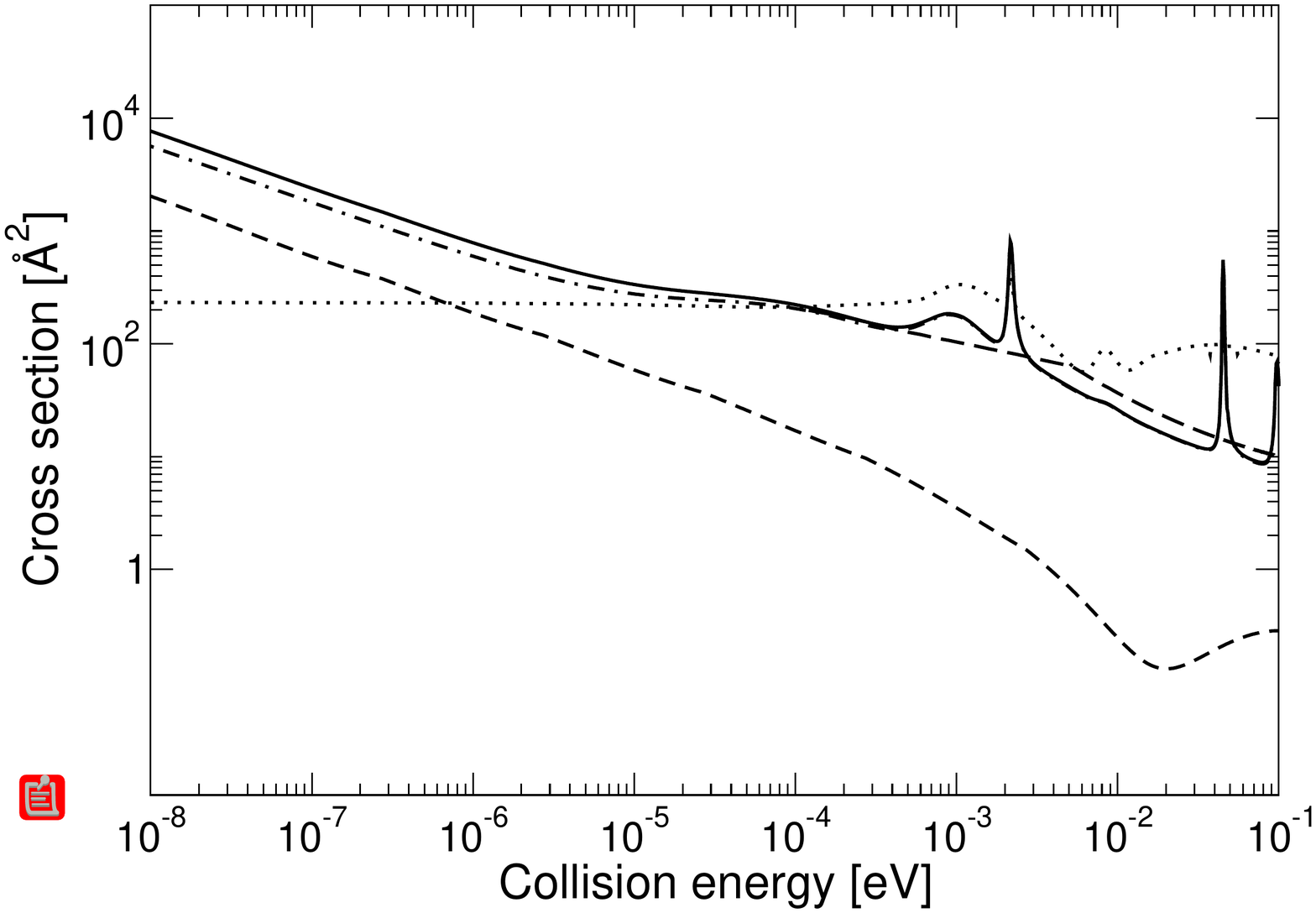}
\caption{Low-energy cross sections for hydrogen--antihydrogen scattering, calculated using the Born-Oppenheimer approximation \cite{Jonsell04}:\emph{solid line}, total inelastic; \emph{dash-dot}, antiproton capture; \emph{short dash}, direct proton--antiproton annihilation; \emph{dotted}, elastic; \emph{long dash}, adiabatic approximation to total inelastic (for energies $> 10^{-3}$~eV).}
\label{fig:HHbarny}
\end{figure}

At very low energies, there are also quantum calculations for the inelastic cross section (including both direct proton-antiproton annihilation and antiproton capture), see figure \ref{fig:HHbarny}. The dominating inelastic channel is rearrangement into Ps($1s$) and Pn($NL$), but surprisingly the dominating Pn state is not  $N=24$, which lies closest to the threshold energy. As was shown in \cite{Armour02}, the dominating final channel has $N=23$. At higher energies, a number of resonances appear, both in the inelastic and the elastic channels. For $E>10^{-3}$ eV I also include the adiabatic approximation to the total inelastic cross section, which, apart from the resonant structure (which the adiabatic method cannot capture), agrees well with the quantum mechanical calculation.

\begin{table}[!h]
\caption{Zero-energy limit of the elastic hydrogen-antihydrogen cross-section, calculated using different methods.}
\label{tab:HaH}
\begin{tabular}{lll}
\hline
$\sigma$ [\AA$^2$] & method & reference \\
\hline
250 & Born-Oppenheimer & \cite{Jonsell01}\\
254 &Kohn variational & \cite{Armour02}\\
208 & Coupled channel & \cite{Sinha03}\\
\hline
\end{tabular}
\vspace*{-4pt}
\end{table}

Figure \ref{fig:HHbarny} also presents quantum mechanical results for the elastic cross section \cite{Jonsell01}.  These results    are based on the Born-Oppenheimer approximation, and are therefore suspect as this approximation breaks down for $R\lesssim R_{\mathrm{FT}}$. Fortunately, there are other calculations of elastic scattering, using methods going beyond the Born-Oppenheimer approximation. In Table \ref{tab:HaH}, I compare the results for the elastic cross section in the zero-energy limit, where the results tend to be very sensitive to the details of the calculation. The agreement is satisfactory.

\subsection{Antihydrogen--helium}

\begin{figure}[!h]
\centering\includegraphics[width=10 cm]{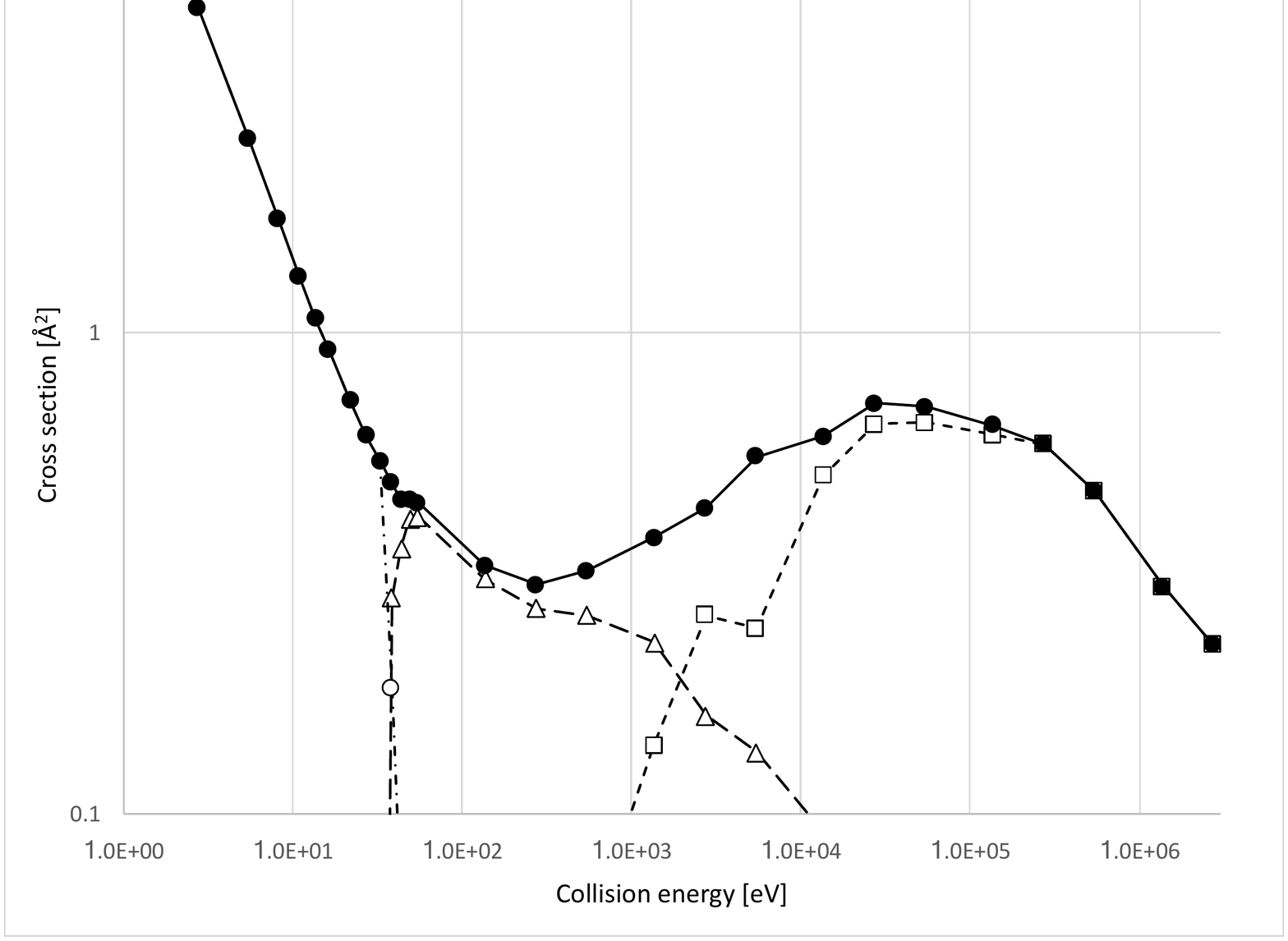}
\caption{Cross sections for helium--antihydrogen scattering, calculated using the FMD method \cite{Cohen06}: \emph{solid line}, total inelastic; \emph{short dash}, ionisation of antihydrogen; \emph{long dash}, $\mathrm{He}^++\bar{p}+\mathrm{Ps}$; \emph{dash-dot}, $(\mathrm{He}^+\bar{p})+\mathrm{Ps}$.}
\label{fig:HeHbarFMD}
\end{figure}

\begin{figure}[!h]
\centering\includegraphics[width=10 cm]{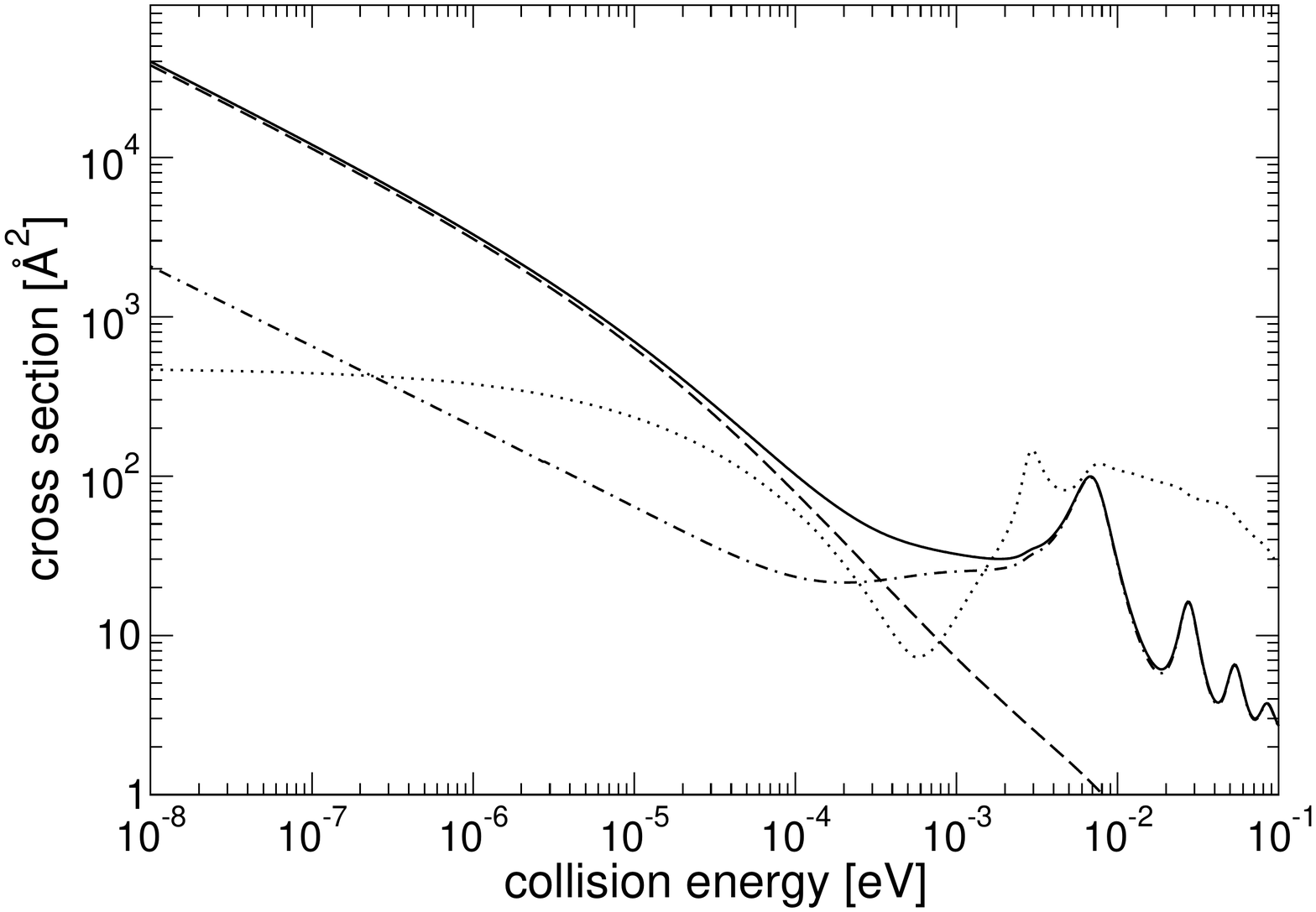}
\caption{Low-energy cross sections for helium--antihydrogen scattering, calculated using the Born-Oppenheimer approximation \cite{Jonsell12}: \emph{solid line}, total inelastic; \emph{dash-dot}; antiproton capture; \emph{short dash}, direct alpha particle--antiproton annihilation; \emph{dotted}, elastic.}
\label{fig:HeHbarLE}
\end{figure}

Turning to helium,  FMD calculations \cite{Cohen06} show that around the ionisation threshold many different charge-transfer and break-up processes compete. The dominating channels are shown figure \ref{fig:HeHbarFMD}.

At lower energies, we again  cannot define the adiabatic cross section, because there is no Fermi-Teller distance (the compound positronium hydride is bound). On the other hand, this makes quantum mechanical approaches based on the Born-Oppenheimer approximation more sound. Such calculations, bases on the very accurate potential calculated by Strasburger \emph{et al.} \cite{Strasburger05}, were performed in \cite{Jonsell12}. In the zero-energy limit, it turns out that the dominating inelastic process is the direct antiproton--alpha particle annihilation \cite{Jonsell04He}, and the second-most important loss process is rearrangement into $\bar{p}\mathrm{He}^+$ and Ps. However, as energy, and hence angular momentum, increases, the rotational barrier shields the nuclei from each other.  Thus the direct annihilation cross section drops rapidly with energy, and above 0.35 meV the rearrangement process dominates.  Both inelastic and elastic cross sections show resonant structure for energies above a few meV (or tens of Kelvin). These resonances are due to pre-dissociative states of the He$\overline{\mathrm{H}}$ compound.

\subsection{Other collisions involving antihydrogen}

Antihydrogen collisions with H$_2$ and H$_2^+$ has also been studied using FMD \cite{Cohen06H2}. Qualitatively, the structure is similar to figures \ref{fig:Hion}, \ref{fig:He} and \ref{fig:HeHbarFMD}, with a sharp increase towards lower energies and different processes taking over at the ionisation thresholds. At lower energies protonium formation is the dominating inelastic channel, while at higher energies ionisation processes dominate. At low energies, protonium formation  is far more likely for $\overline{\mathrm{H}}$--$\mathrm{H}_2$ scattering, compared to $\overline{\mathrm{H}}$--$\mathrm{H}$ scattering. Thus, molecular hydrogen cannot be treated by the simple prescription to multiply the cross section for atomic hydrogen by 2. Unfortunately it is hard to assess the accuracy of these calculations, as there is nothing to compare the results to.

\section{Conclusion}
Most importantly, we have seen that there is an almost  total lack of experimental results at energies below the keV-scale. At lower energies, any experimental information, even at low precision, would be very important to get at least some validation of the various theoretical approaches.

From a theoretical point of view, the high-energy data seems relatively well understood, even if rather laborious calculations in many cases are required to get good quantitative agreement.
We have seen that at energies $\gtrsim 3$~keV, experimental results on antiproton scattering, when available, are well reproduced by theoretical calculations.

At lower energies, the semi-classical approach agrees quite well with the much more complicated full quantum calculations (though available only for hydrogen). If high precision is not required, the FMD method, which is very simple to use, is also an alternative at energies above $\sim 1$eV. At low energies the Langevin and adiabatic methods give results similar to much more complicated methods. 

For antihydrogen, the FMD method has been applied to scattering on the simplest atoms and molecules. At the very low energies, relevant to trapped antihydrogen interacting with cryogenic gases, quantum mechanical methods have been used. Such exist for hydrogen and helium targets, and will perhaps become available for H$_2$. However, targets with more than two electrons would be much more difficult.

\funding{This work was funded by the Swedish Research Council (VR), through grants no.  2008-00442 and 2012-02435.}




\end{document}